\documentclass[a4paper,11pt]{article}
\usepackage{pos}

\usepackage{amsmath}
\usepackage{amsfonts}
\usepackage{physics}
\usepackage{slashed}
\usepackage{simplewick}
\usepackage{siunitx}

\usepackage{graphicx}
\usepackage{overpic}
\usepackage{float}
\usepackage[caption=false]{subfig}
\usepackage{multirow}

\usepackage{tikz}
\usetikzlibrary{calc}
\usetikzlibrary{patterns}
\usetikzlibrary{positioning,decorations.pathmorphing,decorations.markings}
\usetikzlibrary{shapes,arrows}

\usepackage{feynmp-auto}
\usepackage{multirow}
\usepackage{adjustbox}

\usepackage{enumitem}

\usepackage{hyperref}

\usepackage{color}
\usepackage{comment}

\usepackage{xcolor}

\title{Progress in calculation of the fourth Mellin moment of the pion light-cone distribution amplitude using the HOPE method}
\ShortTitle{Fourth Mellin moment of Pion LCDA from the HOPE Method}

\manuallySeparateAuthors
\author[a,b]{William~Detmold,}
\author[a,c]{ Anthony~V.~Grebe,}
\author[d]{ Issaku~Kanamori,}
\author[e,f,g]{ C.-J.~David~Lin,}
\author*[e,f,h,i]{ Robert~J.~Perry}
\author[j]{ and Yong~Zhao}
\author{ for the HOPE collaboration}

\affiliation[a]{Center for Theoretical Physics,\\ Massachusetts Institute of Technology,
Cambridge, MA 02139, USA}
\affiliation[b]{The NSF AI Institute for Artificial Intelligence and Fundamental Interactions}
\affiliation[c]{Theoretical Physics Department, Fermi National Accelerator Laboratory,
P.O. Box 500, Batavia, IL 60410, USA}
\affiliation[d]{RIKEN Center for Computational Science,\\ Kobe 650-0047, Japan}
\affiliation[e]{Institute of Physics,\\ National Yang Ming Chiao Tung University, Hsinchu 30010, Taiwan}
\affiliation[f]{Centre for Theoretical and Computational Physics,\\ National Yang Ming Chiao Tung University, Hsinchu 30010, Taiwan}
\affiliation[g]{Centre for High Energy Physics,\\ Chung-Yuan Christian University, Chung-Li, 32032, Taiwan}
\affiliation[h]{Departament de F\'isica Qu\`antica i Astrof\'isica (FQA),\\ Universitat de Barcelona (UB),  c. Mart\'i Franqu\'es, 1, 08028 Barcelona, Spain}
\affiliation[i]{Institut de Ci\`eincies del Cosmos (ICCUB),\\ Universitat de Barcelona (UB), c. Mart\'i i Franqu\'es, 1, 08028 Barcelona, Spain}
\affiliation[j]{Physics Division,\\ Argonne National Laboratory, Lemont, IL 60439, USA}

\emailAdd{perryrobertjames@gmail.com}

\abstract{
The pion light-cone distribution amplitude (LCDA) is a central non-perturbative object of interest for the calculation of high-energy exclusive processes in quantum chromodynamics. 
This article describes the progress in the lattice QCD calculation of the fourth Mellin moment of the pion LCDA using a heavy-quark operator product expansion (HOPE). 
}

\FullConference{%
The 39th International Symposium on Lattice Field Theory,\\
8th-13th August, 2022,\\
Rheinische Friedrich-Wilhelms-Universität Bonn, Bonn, Germany
}

\begin{document}
\maketitle

\section{Introduction}
\label{sec:intro}
The pion light-cone distribution amplitude (LCDA) is a central non-perturbative object in the description of a range of exclusive processes in high-energy quantum chromodynamics (QCD)~\cite{Lepage:1980fj}. The pion LCDA is denoted $\phi_\pi(\xi,\mu^2)$ and can be defined via the matrix element for the transition amplitude between the vacuum and the (charged) pion state,
\begin{equation}
\bra{0}\overline{\psi}_d(z)\gamma_\mu \gamma_5 \mathcal{W}[z,-z]\psi_u(z)\ket{\pi^+(\mathbf{p})}=if_\pi p_\mu\int_{-1}^{1}d\xi\, e^{-i\xi p\cdot z}\phi_\pi(\xi,\mu^2)\,,
\end{equation}
where $\mathcal{W}[z,-z]$ is a light-like ($z^2=0$) Wilson line connecting $-z$ and $z$ and $\mu$ is the renormalization scale. In the above equation, $f_\pi$ is the pion decay constant and $p^\mu$ is the four-momentum of the pion. In the light-cone gauge, the pion LCDA can be interpreted as the probability amplitude to convert the pion into a state of a quark and an antiquark carrying momentum fractions $(1 + \xi)/2$ and $(1 - \xi)/2$, respectively. 

It is well-known that the non-perturbative matrix elements required for factorization theorems generically contain non-local operators defined with light-like separations. Matrix elements of such operators cannot be directly accessed via a Euclidean field theory. As a result, a number of alternative strategies for extracting information about these non-perturbative matrix elements
using LQCD have been proposed in the last two
decades~\cite{Aglietti:1998ur,Liu:1999ak,Detmold:2005gg,Braun:2007wv,Davoudi:2012ya,Ji:2013dva,Chambers:2017dov,Radyushkin:2017cyf,Ma:2017pxb}. This work follows the method suggested in
Ref.~\cite{Detmold:2005gg} and expanded on in Refs.~\cite{Detmold:2018kwu,Detmold:2020lev,Detmold:2021uru,Detmold:2021qln,Detmold:2021plw}. The method relates hadronic matrix elements directly computable in Euclidean field theory to a heavy-quark operator product expansion (HOPE), where the non-perturbative information about the LCDA is encoded in its Mellin moments that can be determined by fitting lattice data to the HOPE. For this reason, the approach is known as the HOPE method. The use of a fictitious heavy-quark in the computation has the advantage that the quark mass serves as an additional hard scale which suppresses higher-twist corrections, and can be varied to study the residual higher-twist effects present in the numerical data.

In Ref.~\cite{Detmold:2021qln}, the HOPE method was implemented to extract the second Mellin moment of the pion LCDA. The success of this approach motivates the current attempt to extend this formalism to extract the fourth Mellin moment, $\expval{\xi^4}$, for which little is currently known. The only existing determination of the fourth moment from lattice QCD is $\expval{\xi^4}(\mu=2~\si{GeV})=0.124(11)(20)$ at a single lattice spacing of $a=0.076~\si{fm}$~\cite{Gao:2022vyh}. In this proceedings the progress towards the determination of the fourth Mellin moment of the pion LCDA using the HOPE method is reported. The structure of this proceedings is as follows: the HOPE method is briefly reviewed in Sec.~\ref{sec:formalism}, the analysis method is discussed in Sec.~\ref{sec:strategy} and the numerical implementation is explained in Sec.~\ref{sec:lattice_details}. Finally, the conclusions of this work are given in Sec.~\ref{sec:conclusion}. 

\section{The Heavy-Quark Operator Product Expansion}
\label{sec:formalism}

\subsection{Relevant results from the HOPE strategy}
\label{sec:HOPE_strategy}
As referenced above, direct computation of matrix elements of light-like operators is not possible in a Euclidean field theory like lattice QCD. The conventional approach to this problem is to use the operator product expansion (OPE) to expand the light-like operator as an infinite sum of local operators, whose analytic continuation to imaginary time is trivial. However, this approach has a serious drawback: the use of a lattice regulator leads to non-perturbative operator mixing which does not appear in the continuum~\cite{Martinelli:1987si}. These operators are lower dimensional and thus lead to power-divergences. This issue has limited the conventional approach to a determination of the lowest non-trivial (the 2nd) moment of the pion LCDA, where a special choice of kinematics prevents the occurence of such operator mixings~\cite{Martinelli:1987si}. In the approach studied here, this issue is avoided by computing a hadronic matrix element with conserved currents. The use of conserved currents ensures that the continuum limit exists after multiplicative renormalization, avoiding the issue of power divergences. The hadronic matrix element studied here is
\begin{equation}
V_1^{[ \mu\nu ]} (q,p) =\int d^4z\, e^{iq\cdot z}\bra{0}T\{J_\Psi^\mu(z/2)J_\Psi^\nu(-z/2)\}\ket{\pi(\mathbf{p})}\,,
\end{equation}
where the currents $J_\Psi^\mu$ are defined as
\begin{equation}
J_{l,\Psi}^\mu=\overline{\Psi}\gamma^\mu \gamma_5 \psi_l+\overline{\psi}_l\gamma^\mu \gamma_5 \Psi\,,\label{eq:heavy_current}
\end{equation}
where $\psi_l$ is the light-quark ($l=u,d$), $\Psi$ is the heavy-quark and the subscript $1$ has been added to emphasize that this matrix element is due to the contribution from the lowest lying pseudoscalar meson, the pion.

One can compute the HOPE expression for this matrix element~\cite{Detmold:2005gg}. As discussed in Ref.~\cite{Detmold:2021uru}, including perturbative corrections and re-summed target-mass effects, one finds
\begin{equation}
\label{eq:Mellin_OPE_had_amp_target_mass}
V_1^{[ \mu\nu ]} (q,p) = - \frac{2 i \epsilon^{\mu\nu\rho\sigma}
 q_{\rho} p_{\sigma}}{\tilde{Q}^{2}}  f_{\pi} \sum_{n=0,{\mathrm{even}}}^{\infty}
 C_{W}^{(n)} (\tilde{Q}^{2},
 \mu, m_\Psi)  \expval{ \xi^{n} }  \left [ \frac{\zeta^{n} {\mathcal{C}}_{n}^{2}
(\eta)}{2^{n}(n+1)\tilde{Q}^{2}}\right ]\, ,
\end{equation}
where $m_\Psi$ is the heavy-quark mass, $p^\mu$ and $q^\mu$ are the four-momenta of the pion and current, respectively. The scalar functions depend on the kinematic invariants
\begin{align}
\tilde{Q}^2&=Q^2+m_\Psi^2\,,
\\
\zeta&=\frac{\sqrt{p^2q^2}}{\tilde{Q}^2}\,,
\\
\eta&=\frac{p\cdot q}{\sqrt{p^2q^2}}\,.
\end{align}
The Wilson coefficients, $C_{W}^{(n)}$, have been computed in the $\overline{\text{MS}}$ scheme~\cite{Detmold:2021uru}, and thus the resulting heavy-quark and Mellin moments extracted from this expansion are also to be understood in this scheme at the renormalization scale $\mu$. Performing a Fourier transform in the temporal direction, one obtains
\begin{equation}
R_1^{[\mu\nu]}(t,\mathbf{p},\mathbf{q})=\int \frac{dq_4}{(2\pi)}\, e^{-iq_4 t} V_1^{[ \mu\nu ]} (q,p)\label{eq:tmr}
\end{equation}
This quantity is directly computable in a Euclidean field theory.

\section{Analysis Strategy: Introducing the Ratio Method}
\label{sec:strategy}
The starting point to compute Eq.~\eqref{eq:tmr} is the three-point correlator
\begin{equation}
C_3^{[\mu\nu]}(t_e,t_m,\mathbf{p},\mathbf{q})=\int d^3 x_e\, d^3x_m\, e^{-i\mathbf{p}_e\cdot \mathbf{x}_e}e^{-i\mathbf{p}_m\cdot \mathbf{x}_m}\bra{0}T\{J^{[\mu}(x_e)J^{\nu]}(x_m)\mathcal{O}_\pi(0)\}\ket{0}\,,\label{eq:three_point}
\end{equation}
where $\mathbf{p}=\mathbf{p}_e+\mathbf{p}_m$ and $\mathbf{q}=(\mathbf{p}_e-\mathbf{p}_m)/2$ are the pion and current three-momentum, $\mathcal{O}_\pi$ is the pseudoscalar operator described below, and $J_\Psi$ is chosen as in Eq.~\eqref{eq:heavy_current}. Inserting a complete set of states between the pseudoscalar interpolating operator and one of the currents leads to
\begin{equation}
C_3^{[\mu\nu]}(t_e,t_m,\mathbf{p},\mathbf{q})=\frac{Z_1}{2E_1}e^{-E_1(t_e+t_m)/2}R_1^{\mu\nu}(t_e-t_m,\mathbf{p},\mathbf{q})+\frac{Z_2}{2E_2}e^{-E_2(t_e+t_m)/2}R_2^{\mu\nu}(t_e-t_m,\mathbf{p},\mathbf{q})+\dots\,,
\end{equation}
where $R_1^{\mu\nu}$ is the hadronic matrix element of interest described in Eq.~\eqref{eq:tmr}. In order to extract $R_1^{\mu\nu}$ from lattice data, which in general contains additional terms from excited-state contamination, previous studies~\cite{Detmold:2018kwu,Detmold:2021uru,Detmold:2021plw,Detmold:2021qln} compute the above three-point correlator at large Euclidean time, and then construct the ratio
\begin{equation}
\frac{C_3^{[\mu\nu]}(t_e,t_m,\mathbf{p},\mathbf{q})}{\frac{Z_1}{2E_1}e^{-E_1(t_e+t_m)/2}}\to R_1^{\mu\nu}(t_e-t_m,\mathbf{p},\mathbf{q})\,,
\end{equation}
which asymptotically approaches the matrix element of interest in the limit of large $(t_e+t_m)$. Such a ratio requires a precise analysis of two-point correlators to extract the desired energies and overlap factor, $Z_{1}$. In this analysis, an alternative approach to analyzing the numerical data is explored. Rather than constructing a ratio of three-point to extracted two-point parameters, $E_{1}$ and $Z_{1}$, in this analysis a non-trivial ratio of three-point correlators is constructed. The advantages of this approach are that one no longer needs a precise determination of quantities from a study of two-point correlator data, and one no longer needs to renormalize the current, since the renormalization factor cancels in the ratio.

To begin, the above expression may be simplified by the variable redefinitions
\begin{align}
t_+=t_e+t_m,
\\
t_-=t_e-t_m.
\end{align}
One then constructs the ratio of three-point correlators:
\begin{equation}
\begin{split}
\mathcal{R}(t_+,t_-;\pm \delta)&=\frac{C_3^{[\mu\nu]}(t_e\pm \delta a,t_m \mp \delta a,\mathbf{p},\mathbf{q}))}{C_3^{[\mu\nu]}(t_e,t_m,\mathbf{p},\mathbf{q}))}=\frac{\mathcal{C}_3^{[\mu\nu]}(t_+,t_-\pm 2\delta a,\mathbf{p},\mathbf{q}))}{\mathcal{C}_3^{[\mu\nu]}(t_+,t_-,\mathbf{p},\mathbf{q}))}
\\
&=\frac{R_1^{[\mu\nu]} (t_-\pm 2\delta a,\mathbf{p},\mathbf{q})[1+A(t_-\pm2\delta a)e^{-\Delta Et_+/2}+\dots]}{R_1^{[\mu\nu]} (t_-,\mathbf{p},\mathbf{q})[1+A(t_-)e^{-\Delta Et_+/2}+\dots]}\,,
\end{split}
\end{equation}
where
\begin{equation}
A(t_-,\mathbf{p},\mathbf{q})=\frac{Z_2 E_1 R_2^{[\mu\nu]}(t_-,\mathbf{p},\mathbf{q})}{Z_1 E_2 R_1^{[\mu\nu]}(t_-,\mathbf{p},\mathbf{q})}\,.
\end{equation}
At large Euclidean times, the combination $A(t_-,\mathbf{p},\mathbf{q})e^{-\Delta Et_+/2}$ is expected to be small, and so the denominator may be expanded to obtain
\begin{equation}
\mathcal{R}(t_+,t_-;\pm \delta )=\frac{R_1^{[\mu\nu]} (t_-\pm \delta a,\mathbf{p},\mathbf{q})}{R_1^{[\mu\nu]} (t_-,\mathbf{p},\mathbf{q})}[1+(A(t_-\pm \delta a,\mathbf{p},\mathbf{q})-A(t_-,\mathbf{p},\mathbf{q}))e^{-\Delta Et_+/2}+\dots]\,.\label{eq:ratio}
\end{equation}
To check the validity of this, one can plot $\mathcal{R}$ for a range of $t_+$ and study the plateau. Excited state contamination is suppressed in the limit of large $t_+$. In this study $\delta=-1$ is chosen.

\section{Numerical Implementation}
\label{sec:lattice_details}

The gauge fields used in this study were tuned to a constant physical volume of $L=1.92~\si{fm}$ and a constant pion mass of $m_\pi\sim 0.55~\si{GeV}$. Leading finite volume effects arise from the `around-the-world' pion contributions, which are small at this pion mass ($\exp(-m_\pi L)\approx 1\%$) and currently neglected in the analysis. The heavy quark masses were chosen to give approximately constant masses of the heavy-heavy pseudoscalar meson across the two lattices. Further details on the lattice action, including the order-$a$ improvement obtained from the use of Wilson-clover fermions can be found in Ref.~\cite{Detmold:2021qln}.
Further details on the lattices and quark masses used are listed in Table~\ref{tab:lattice_details}.
The required two- and three-point functions were generated using the software package \textsc{Chroma} with the \textsc{QPhiX} inverters~\cite{chroma, qphix}. 

\begin{table}
\centering
  \begin{tabular}{ l l l l l l l }
  \hline\hline
$(L/a)^3 \times (T/a)$ & $a$ (fm) & $N_\text{cfg}$ & $\kappa_l$  & $\kappa_h$ & $m_\Psi^{\overline{\text{MS}}}$ (\si{GeV}) \\ \hline
\multirow{2}{*}{$24^3 \times 48$} & \multirow{2}{*}{0.0813}  & \multirow{2}{*}{6500} & \multirow{2}{*}{0.1349}   
& 0.120 & $2.0~\si{GeV}$ \\ 
 &  &  &  & 0.110 & $2.6~\si{GeV}$ \\ 
 \hline
\multirow{3}{*}{$32^3 \times 64$ } & \multirow{3}{*}{0.0600}  & \multirow{3}{*}{4500} & \multirow{3}{*}{0.1352} & 0.125 & $2.0~\si{GeV}$ \\  
 &  &  &  &  0.118 & $2.6~\si{GeV}$ \\
 &  &  &  & 0.110 & $3.4~\si{GeV}$  \\
 \hline
  \end{tabular}
  \caption{Details of the gauge field configurations and quark masses used in this study. These configurations were generated in Ref.~\cite{Detmold:2018zgk}. Heavy quark masses obtained from the fit of numerical data to the one-loop HOPE are also given. }\label{tab:lattice_details}
\end{table}

\subsection{Reducing Excited State Contamination}
As discussed previously in Refs.~\cite{Detmold:2020lev,Detmold:2021qln}, in order to extract information about the higher Mellin moments, one must use large hadronic momentum $\mathbf{p}=2\pi\mathbf{n}/L$. In this study, equivalent momenta with $\mathbf{n}^2=4$ were chosen. It is well-known that this requirement of larger hadronic momentum makes the isolation of the ground state more difficult, since the energy gap $\Delta E(\mathbf{p})=E_2(\mathbf{p})-E_1(\mathbf{p})$ shrinks as $|\mathbf{p}|$ increases, and thus the contamination to the ground state persists to larger Euclidean times. Furthermore, statistical degradation of the signal at large Euclidean times makes extraction of information from the numerical data difficult. While the mass-gap is fixed for a given choice of lattice parameters, it is possible to change the overlap factors $Z_i$ by optimizing the pseudoscalar interpolating operator $\mathcal{O}_\pi(x)$. In this study operator smearing and the variational method are combined to produce an interpolating operator with greater overlap with the ground state pion. 

\begin{figure}
\centering
\includegraphics[scale=0.45]{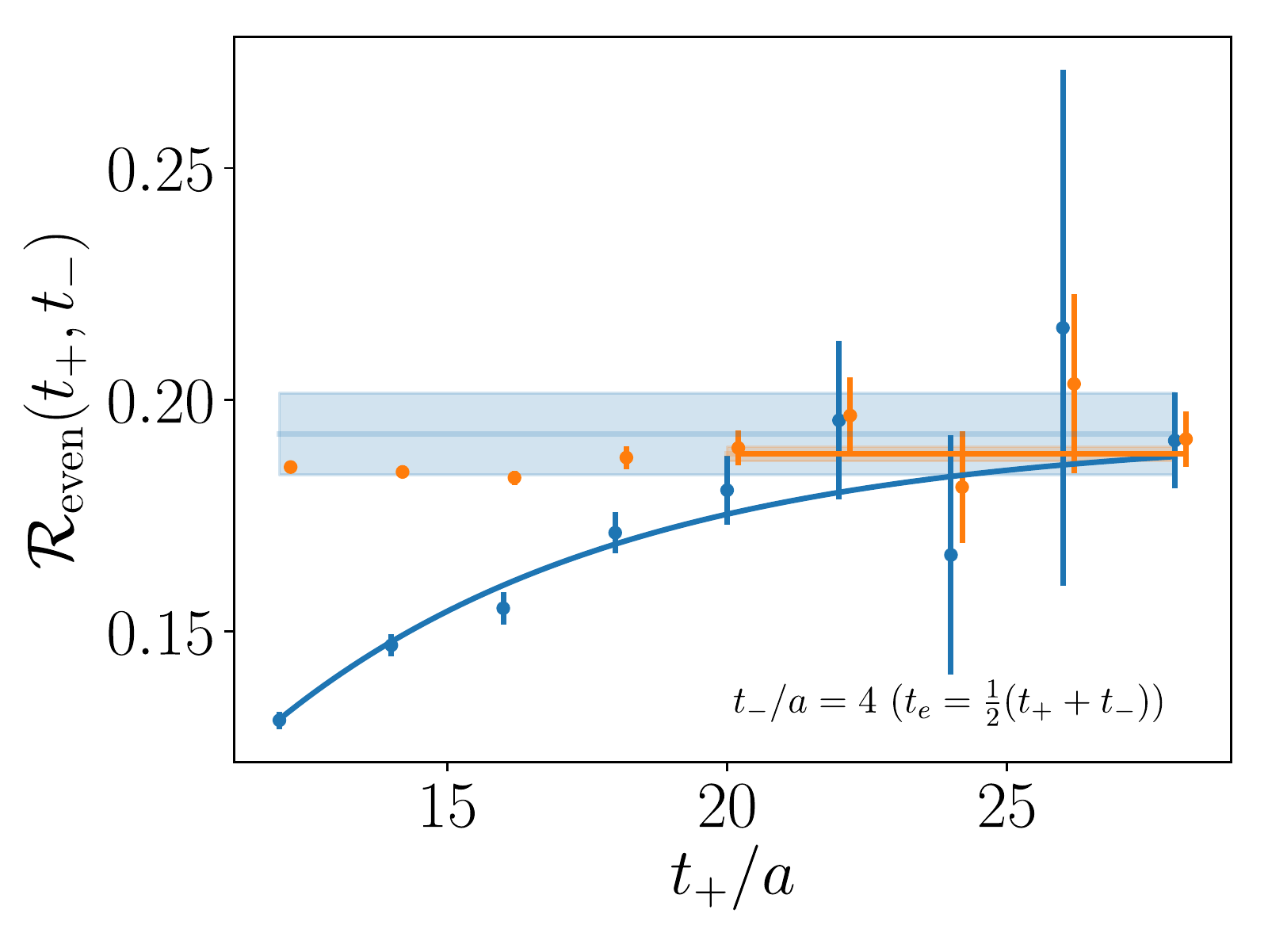}
\includegraphics[scale=0.45]{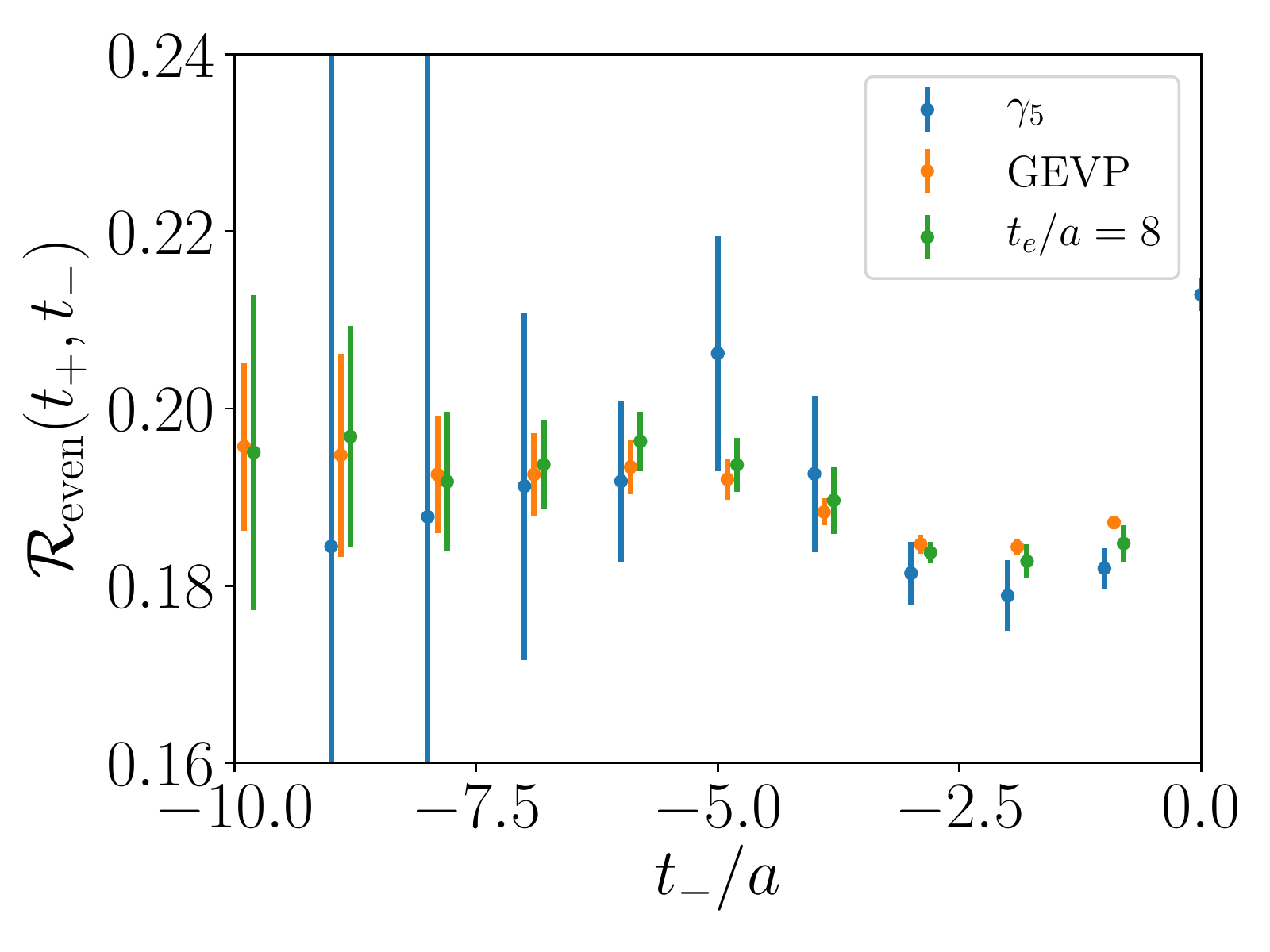}
\vspace{-10pt}
\caption{Studying the effect of operator optimization on the excited state contamination. The left plot shows the excited state dependence for fixed $t_-/a=4$. Data in blue arises from $\mathcal{O}^1$ alone, while data in orange is the GEVP-optimized ratio. Both datasets are extrapolated to the limit of infinite $t_+/a$, and the resulting shaded bands show the extrapolated results from the fit. Note that data for the GEVP-optimized ratio agrees with the extrapolated results at earlier $t_+/a$. The right plot compares the extrapolated ratio constructed from $\mathcal{O}^1$ alone (denoted $\gamma_5$) and the GEVP-optimized ratio (denoted GEVP) with the ratio computed at fixed $t_e/a=8$ using the GEVP-optimized ratio. The agreement within statistical errors between the three datasets implies that at $t_e/a=8$, residual excited state contamination is negligible with respect to statistical errors, and thus an extrapolation to $t_+/a\to\infty$ is unnecessary.}\label{fig:operator_optimization}
\end{figure}

In particular, this study utilizes a combination of gauge-invariant Gaussian smearing~\cite{Gusken:1989qx} and momentum smearing~\cite{Bali:2016lva} to improve the spatial overlap of the local quark bilinear operator with the physical pion state. The width of the Gaussian smearing was taken as the inverse pion mass, $a\omega_\text{smear}=\{4.5, 6.0\}$ for $L/a=\{24,32\}$, and a smearing momentum fraction of $\zeta=0.8$, as proposed in Ref.~\cite{Bali:2016lva} was employed. While the improvement over just Gaussian smearing was clear, further improvement of the operator overlap was obtained  by considering a larger set of interpolating operators. By increasing the operator basis, one can construct an improved variational estimate of the ground state operator~\cite{Michael:1982gb,Luscher:1990ck}. This study used a two-dimensional operator basis given by
\begin{align}
\mathcal{O}^1(x)&=\overline{\psi}_d(x)\gamma_5\psi_u(x)\,,
\\
\mathcal{O}^2(x)&=\overline{\psi}_d(x)\gamma_4\gamma_5\psi_u(x)\,.
\end{align}
In order to construct the optimized operator, one calculates the two-point correlator for all combinations of the basis operators at the source and sink:
\begin{equation}
C^{ij}(t,\mathbf{p})=\int d^3x\, e^{i\mathbf{p}\cdot \mathbf{x}} \bra{0}T\{\mathcal{O}^i(x)\mathcal{O}^{j\dagger}(0)\}\ket{0}\,.
\end{equation}
One then solves the generalized eigenvalue problem (GEVP)~\cite{Luscher:1990ck}:
\begin{equation}
C^{ij}(t,\mathbf{p})v_n^j=\lambda_nC^{ij}(t_0,\mathbf{p})v_n^j\,,
\end{equation}
where $\lambda_n=e^{-E_n(t-t_0)}$ are the eigenvalues and $v_n^i$ are the elements of the corresponding eigenvector.
An optimized interpolating operator for the ground state ($n=1$) may be constructed by writing
\begin{equation}
\mathcal{O}_\pi(x)=v_1^i\mathcal{O}^i(x)\,,
\end{equation}
and the optimized three-point correlator is constructed from this linear combination. The reduction in excited state contamination from this procedure is shown in Fig.~\ref{fig:operator_optimization}. As can be seen, an impressive reduction in the excited-state contamination can be observed from the use of an enlarged basis of interpolating operators. 

In order to extract the matrix element of interest, one would ideally like to compute the ratio defined above at a range of $t_+/a$ and then extrapolate to the $t_+\to\infty$ limit. Due to the use of the sequential source method in the construction of the three-point correlator (see Ref.~\cite{Detmold:2021qln} for more details), additional $t_+$ values require a linear increase in runtime. Instead, it was noted that with the use of the improved variational estimate (denoted GEVP in Fig.~\ref{fig:operator_optimization}), data at fixed $t_e/a$ was consistent with the $t_+\to\infty$ extrapolated single operator and variational estimate data within statistical errors. As a result, data was computed at fixed $t_e\sim\{0.56~\si{fm},0.64~\si{fm}\}$ and used to construct the ratio analysed in the following section.

\subsection{Data Analysis}
\begin{figure}
\centering
\includegraphics[scale=0.45]{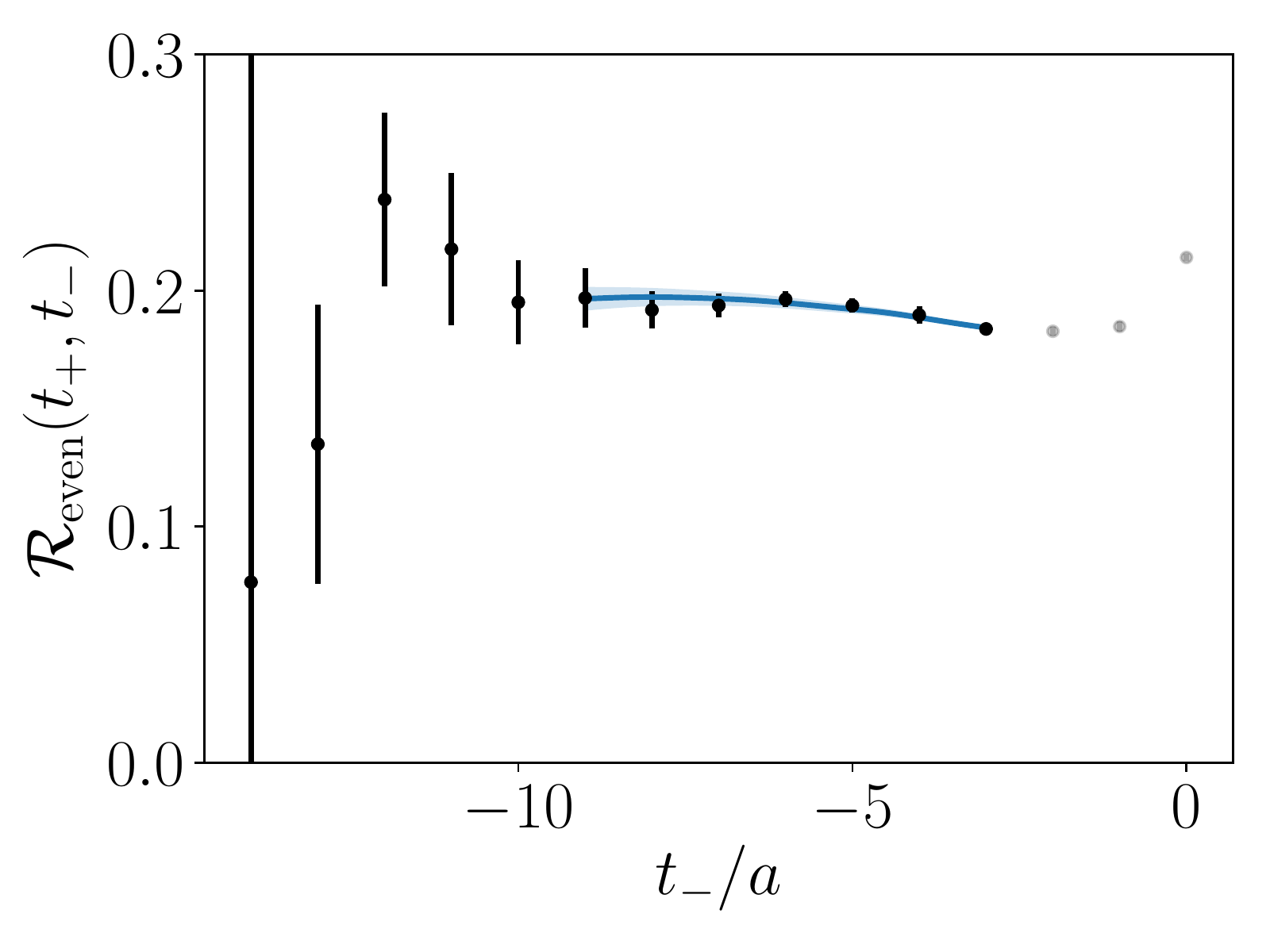}
\includegraphics[scale=0.45]{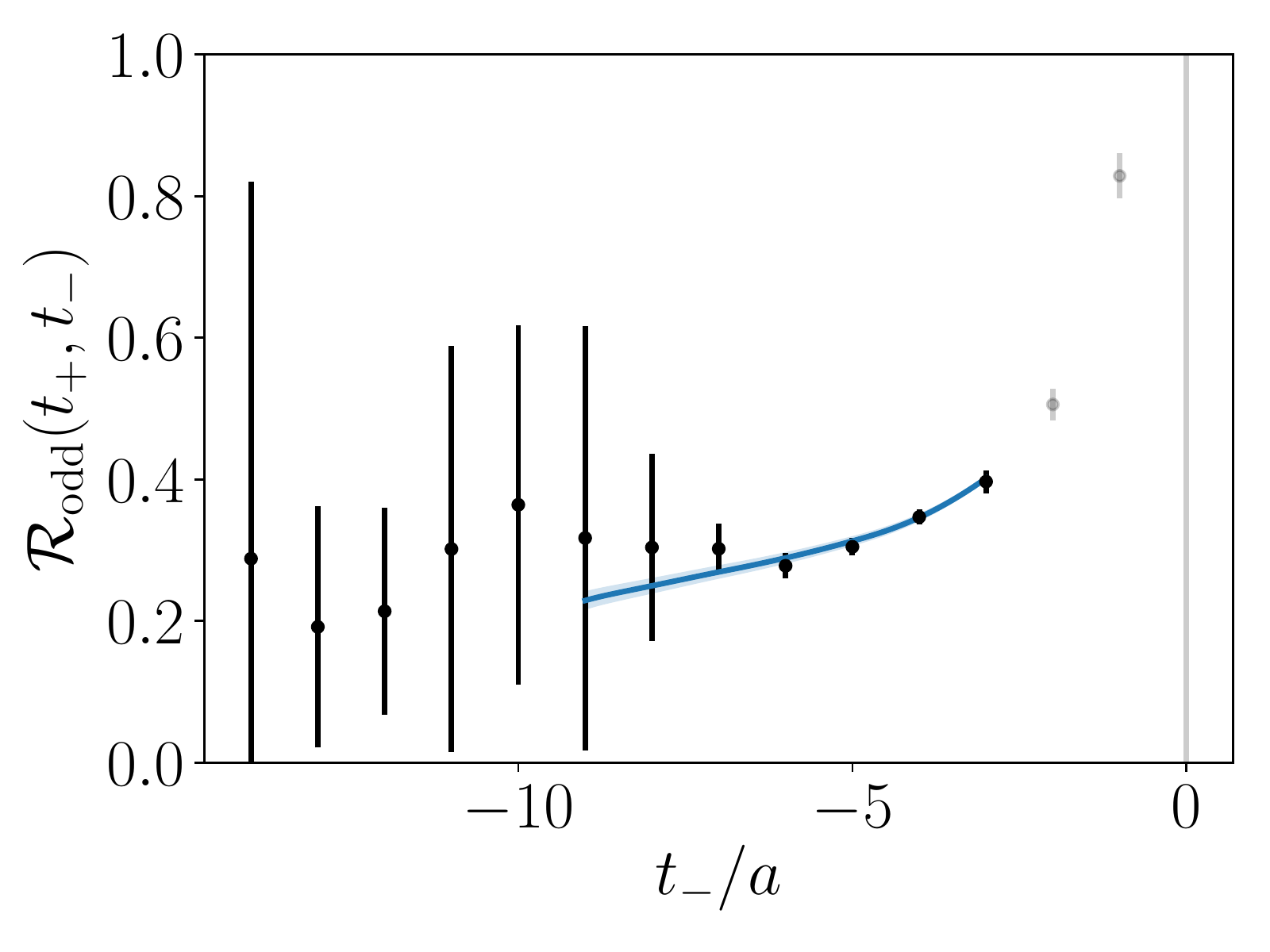}
\vspace{-10pt}
\caption{Characteristic fit of numerical data to continuum HOPE expression at $L/a=24$. The resulting values for fit parameters were $m_\Psi=2.0~\si{GeV}$, $\expval{\xi^2}=0.17 \pm 0.04$, $\expval{\xi^4}=0.07 \pm 0.02$ at a renormalization scale of $\mu=2.0~\si{GeV}$. The quoted uncertainties are purely statistical.}
\label{fig:numerical_data}
\end{figure}

After constructing the ratio described in the previous sections using the optimized interpolating operator, a global fit was performed to the ratio of symmetric in $t_-$ (even) and antisymmetric in $t_-$ (odd) pieces. A characteristic fit for one dataset is shown in Fig.~\ref{fig:numerical_data}. As previously explained in Ref.~\cite{Detmold:2021qln}, one expects uncontrolled lattice artifacts at distances smaller than three units in lattice time, ie, for $t_-/a\leq2$. Thus data are fit from $t_-/a=3$ onwards. The one-loop form of the HOPE expression~\cite{Detmold:2021uru} requires a fit of the heavy-quark mass $m_\Psi$ as well as the second and fourth Mellin moments.

\begin{figure}
\centering
\includegraphics[scale=0.45]{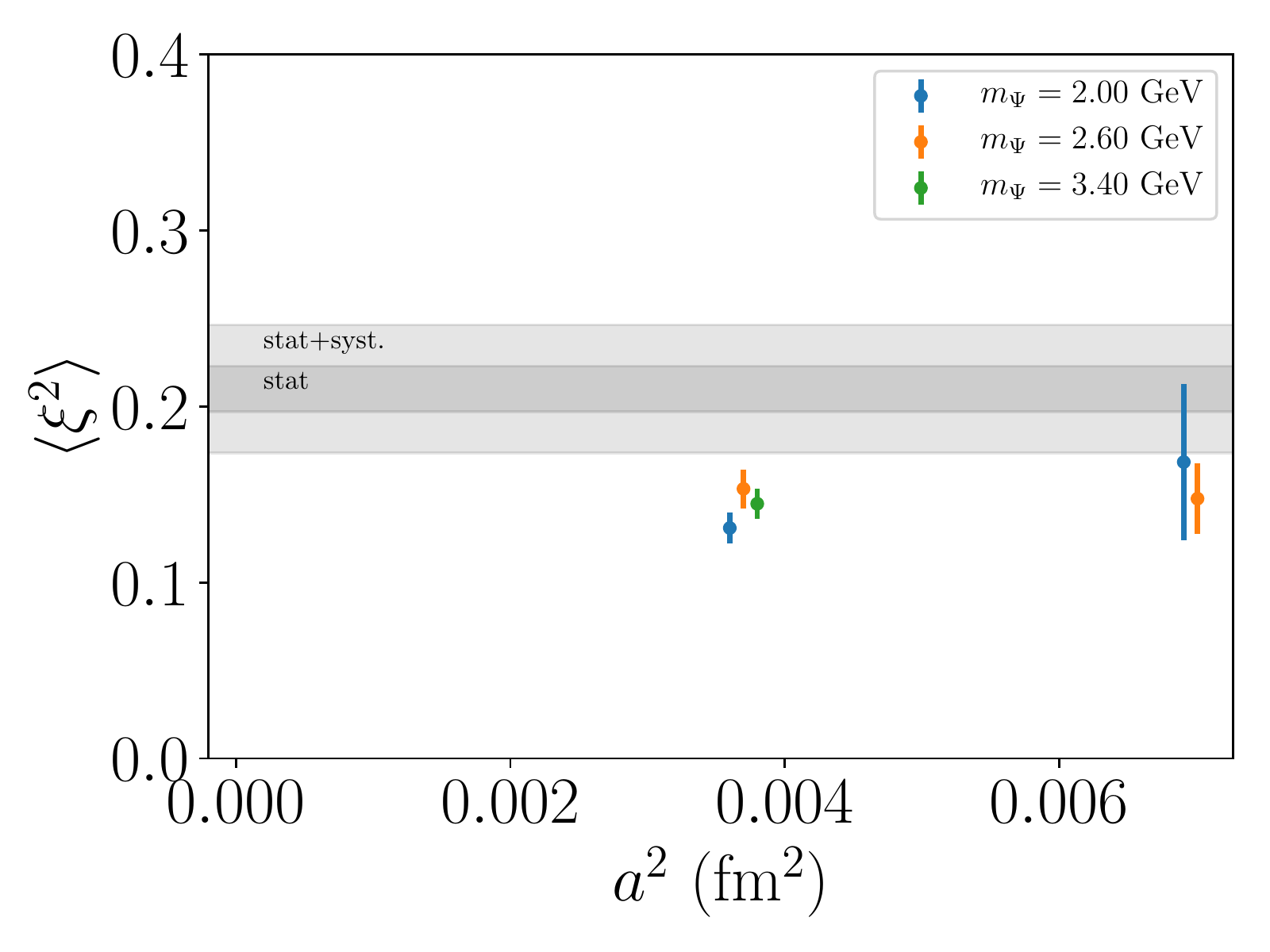}
\includegraphics[scale=0.45]{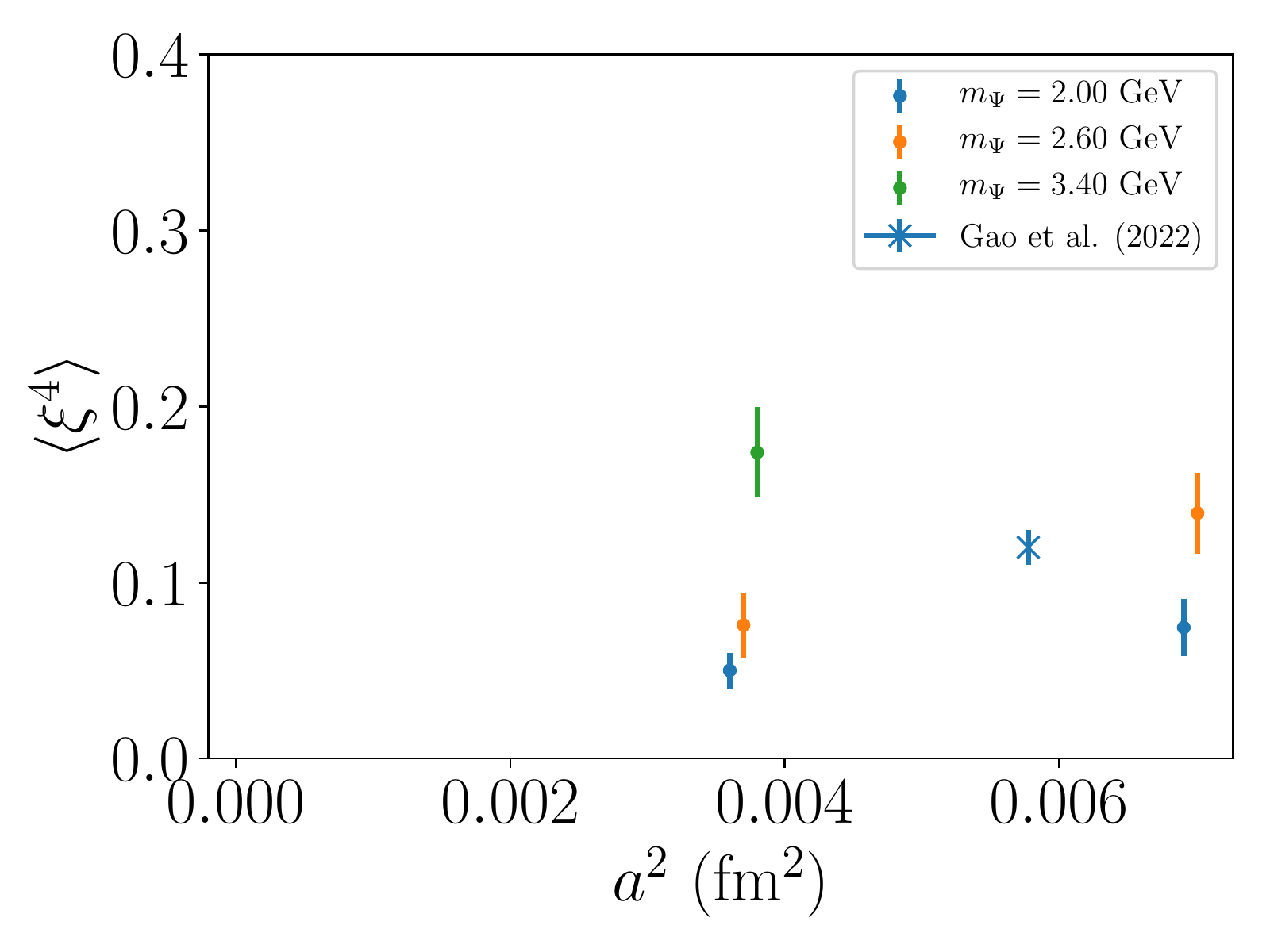}
\vspace{-10pt}
\caption{Comparison of current numerical results for the second and fourth Mellin moments. In Ref.~\cite{Detmold:2021qln}, the second Mellin moment was computed using a superset of the configurations used here. The result in the continuum, twist-two limit is shown by the grey band. While less is known about the fourth Mellin moment of the LCDA, there exists one publication at one lattice spacing which suggests a value of $\expval{\xi^4}=0.124(11)(20)$~\cite{Gao:2022vyh}, which is broadly speaking compatible with the determinations obtained in this work.}
\label{fig:xin}
\end{figure}

Since data at finite lattice spacing are fit to a continuum heavy quark OPE, one expects that the fit parameters contain residual higher-twist and lattice artifacts. These artifacts may be examined in Fig.~\ref{fig:xin}. Unfortunately, although early results are promising, with just two lattice spacings it is not possible to perform a rigorous extrapolation to the continuum limit. With more data, a continuum, twist-two extrapolation of the form
\begin{equation}
\expval{\xi^n}(a,m_\Psi)=\expval{\xi^n}+\frac{A}{m_\Psi}+Ba^2+Ca^2m_\Psi+Da^2m_\Psi^2
\end{equation}
will become possible. To demonstrate that the data are reasonable, the continuum, twist-two second Mellin moment computed using a superset of the gauge ensembles (see Ref.~\cite{Detmold:2021uru}) used in this study is represented as a grey band. It is also interesting to note that the only other calculation of the fourth Mellin moment (Ref.~\cite{Gao:2022vyh}), calculated at a single lattice spacing at the physical pion mass using dynamical quarks, predicts a result which is broadly consistent with the determinations from the HOPE approach at comparable lattice spacings. 

\section{Conclusion}
\label{sec:conclusion}
In this article, the progress towards a continuum limit determination of the fourth Mellin moment of the pion LCDA at a pion mass of $m_\pi\sim 0.55~\si{GeV}$ was discussed. A new strategy for analyzing the numerical data which relied primarily on three-point data was presented. Information about the higher Mellin moments is increasingly kinematically suppressed by the magnitude of the hadron momentum. Thus it is necessary to study the hadronic matrix elements at larger hadronic momentum leading in turn to increased excited-state contamination. A combination of momentum smearing and a variational basis was used to optimize the pseudoscalar interpolating operator to reduce excited-state contamination. Data for the aforementioned ratio, ${\mathcal{R}}$ as defiend in Eq.~\eqref{eq:ratio},  were presented, and the heavy-quark mass and second and fourth Mellin moments were extracted from the numerical data at finite lattice spacing. While the current dataset (two lattice spacings and up to three heavy quarks) was insufficient for a reliable combined continuum, twist-two extrapolation, the numerical determinations of the second Mellin moment appear to agree relatively well with the previous determination of this quantity using the same gauge fields, and the fourth Mellin moment also appears to be in reasonable agreement with the single dynamical determination of this quantity at finite lattice spacing~\cite{Gao:2022vyh}. 

\section*{Acknowledgements}
The authors thank ASRock Rack Inc.~for their support of the construction of an
Intel Knights Landing cluster at National Yang Ming Chiao Tung
University, where the numerical calculations were performed.  Help
from Balint Joo in tuning Chroma is acknowledged.
We thank M. Endres for providing the ensembles of gauge field configurations used in this work.
  CJDL and RJP are supported by the Taiwanese NSTC Grant
No.~109-2112-M-009-006-MY3 and NSTC Grant No.~109-2811-M-009-516. RJP has also been supported by project PID2020-118758GB-I00, financed by the Spanish MCIN/ AEI/10.13039/501100011033/.
The work of IK is partially supported by the MEXT as ``Program for
Promoting Researches on the Supercomputer Fugaku'' (Simulation for
basic science: from fundamental laws of particles to creation of
nuclei) and JICFuS.
YZ is supported by the U.S.~Department of Energy, Office of Science, Office of Nuclear Physics, contract no.~DEAC02-06CH11357.
WD and AVG acknowledge support from the U.S.~Department of Energy (DOE) grant DE-SC0011090.
WD is supported by the SciDAC5 award DE-SC0023116. 
WD is also supported in part by the National Science Foundation
under Cooperative Agreement PHY-2019786 (The NSF AI Institute for Artificial Intelligence and Fundamental Interactions, http://iaifi.org/).
This document was prepared by the HOPE Collaboration using the resources of the Fermi National Accelerator Laboratory (Fermilab), a U.S. Department of Energy, Office of Science, HEP User Facility. Fermilab is managed by Fermi Research Alliance, LLC (FRA), acting under Contract No. DE-AC02-07CH11359.

\addcontentsline{toc}{chapter}{Bibliography} 
\bibliographystyle{jhep.bst} 
\bibliography{bibliography} 

\end{document}